\documentclass{emulateapj}
\usepackage{apjfonts}

\newcommand{\source}{\hbox{3C\,442A}}

\newcommand{\nseven}{\hbox{NGC\,7237}}
\newcommand{\nsix}{\hbox{NGC\,7236}}
\newcommand{\chandra}{\textit{Chandra}}
\newcommand{\spitzer}{\textit{Spitzer}}
\newcommand{\rosat}{\hbox{ROSAT}}

\begin{document}

\slugcomment{Feb 13th 2007: Accepted for publication in ApJ Letters}

\title{The effect of a {\it Chandra}-measured merger-related gas
component on the lobes of a dead radio galaxy}

\author{D.M.
Worrall\altaffilmark{1}, M. Birkinshaw\altaffilmark{1},
R.P. Kraft\altaffilmark{2}, M.J. Hardcastle\altaffilmark{3}}
\altaffiltext{1}{Department of Physics, University of Bristol, Royal
Fort, Tyndall Avenue, Bristol BS8 1TL, UK}
\altaffiltext{2}{Harvard-Smithsonian Center for Astrophysics, 60
Garden Street, Cambridge, MA 02138}
\altaffiltext{3}{School of Physics, Astronomy \& Mathematics,
University of Hertfordshire, College Lane, Hatfield AL10 9AB, UK}

%\email{d.worrall@bristol.ac.uk}
%\email{mark.birkinshaw@bristol.ac.uk}
%\email{rkraft@head.cfa.harvard.edu}
%\email{mjh@star.herts.ac.uk}

\begin{abstract}
We use \chandra\ data to infer that an X-ray bright component of gas
is in the process of separating the radio lobes of \source.  This is
the first radio galaxy with convincing evidence that central gas,
overpressured with respect to the lobe plasma and not simply a static
atmosphere, is having a major dynamical effect on the radio structure.
We speculate that the expansion of the gas also re-excites electrons
in the lobes of \source\ through compression and adiabatic heating.
Two features of \source\ contribute to its dynamical state.  Firstly,
the radio source is no longer being powered by a detected active jet,
so that the dynamical state of the radio plasma is
at the mercy of the ambient medium.  Secondly the two early-type
galaxies, \nsix\ and \nseven, one of which was the original host of
\source, are undergoing a merger and have already experienced a close
encounter, suggesting that the X-ray bright gas is mostly the heated combined
galaxy atmospheres.  The lobes have been swept
apart for $\sim 10^8$~yrs by the pressure-driven expansion of the
X-ray bright inner gas.
\end{abstract}

\keywords{galaxies: active ---
galaxies: individual (\objectname{NGC 7237, 3C 442A}) --- galaxies: interactions
--- radio continuuum: galaxies--- X-rays: galaxies}

\section{Introduction}

Radio galaxies have a profound influence on the intergalactic,
intragroup, or intracluster X-ray emitting gas in which they are in
contact, in particular through moving gas via the creation of cavities
and shocks \citep[see the review of][]{jones07}.  In turn the dynamics
of the radio plasma are affected by the inertia of the gas that it
encounters, and it is slowed and redirected as a result.  In this
letter we report the first case where, instead of radio plasma moving
X-ray emitting gas, the X-ray emitting material is exerting a dominant
influence on old radio plasma, pushing apart the lobes of \source.

\citet{birk81} first reported on the absence of jet emission in the
amorphous radio lobes of \source, and this is confirmed by the more
extensive study of \citet{comins91}.  \source\ can therefore be
characterized as a dead radio galaxy, 
in the sense that it contains no radio jet at the level seen in
actively jet-driven radio galaxies of similar power.

A pair of similar mass early-type galaxies, \nsix\ and \nseven, lie
between the radio lobes of \source.  \citet{borneh88} provide
strong evidence that these two galaxies, which are a part of group
containing also a fainter member, are undergoing a merger, and an
interaction model for the binary system is fitted to the data by
\citet{borne88}.

Our \rosat\ HRI observation of \source\ showed extended emission
elongated in the NS direction and lying between the two lobes
\citep{hw99}, but the relatively low sensitivity of the observation
and the lack of X-ray spectral information prevented the physical
relationship between the gas and radio plasma from being addressed.
This has been corrected through a deep \chandra\ observation of the
source.  In this letter we show that X-ray gas is pressing on and
separating the lobes of the radio galaxy.  A more complete discussion
of the X-ray gas, including that surrounding the lobes and the
filamentary small scale structure, is presented elsewhere
\citep{hard07}.

The average recession velocity of the galaxy pair \nseven\ and
\nsix\ \citep{borneh88} leads to a redshift of $z=0.0272$ for \source. We
adopt values for the cosmological parameters of $H_0= 70$~km s$^{-1}$
Mpc$^{-1}$, $\Omega_{\rm {m0}} = 0.3$, and $\Omega_{\Lambda 0} = 0.7$.
Thus \source\ is at a luminosity distance of 119~Mpc, and
1~arcsec corresponds to 546~pc at the source.
J2000 coordinates are used throughout.  Uncertainties are $1\sigma$
for one interesting parameter unless otherwise stated.

\section{Observations}

\begin{deluxetable}{ccccc}
\tablewidth{0pt}
\tablecaption{\chandra\ Observations
\label{tab:chandraobs}}
\tablehead{
\colhead{OBSID} & \colhead{Date} &
\colhead{Duration (ks)} & \colhead{CCD Chip\tablenotemark{a}} & 
\colhead{shift (arcsec)\tablenotemark{b}}
}
\startdata
5635 & Jul 27, 2005 & 27.006 & I0 & 0.495 \\
6353 & Jul 28, 2005 & 13.985 & I0 & 0.594 \\
6359 & Oct  7, 2005 & 19.883 & I2 & 0.373 \\
6392 & Jan 12-13, 2006 & 32.694& I0 & 0.151 
\enddata
\tablenotetext{a}{\nseven\ on this chip.}
\tablenotetext{b}{Astrometric correction to align X-ray nucleus to
the radio core in \nseven, mostly in declination.}
\end{deluxetable}

We observed \source\ in VFAINT data mode with a front-illuminated CCD
chip of the Advanced CCD Imaging Spectrometer (ACIS) on board
\chandra.  Chips I0, I1, I2, I3 and S2 were turned on for the
observation which was broken into four intervals
(Table~\ref{tab:chandraobs}) and taken in full-frame mode with a
readout time of 3.14 seconds.
The data have been re-processed with random pixelization removed
following the software `threads' (http://cxc.harvard.edu/ciao) from
the \chandra\ X-ray Center (CXC).  We used {\sc vfaint} cleaning and
the recommended procedure (M.~Markevitch 2006, ACIS background
cookbook available from http://cxc.harvard.edu)
of including events with status flag 5 set as bad, to
preserve flux close to the CCD node boundaries and other bad pixels.
This is particularly important for the third observation where
\nseven\ lies on bad pixels.  Only events with grades 0,2,3,4,6 are
used in our analysis.  Results presented here use {\sc ciao v3.3.0.1}
and the {\sc caldb v3.2.3} calibration database.

The background count rate was steady through the four observations.
Small astrometric corrections consistent with \chandra's absolute
aspect uncertainties were made to each data set to register the X-ray
nucleus of \nseven\ to the radio core whose position measured from
archival 4.8-GHz VLA B-array data using standard AIPS
procedures is 22$^h$ 14$^m$ 46$^s.880\pm 0.002$, $+13^\circ$ $50'$
$27''.23\pm 0.02$.  The dates, durations, CCD chips in which \nseven\
was contained, and sizes of the astrometric corrections for each
observation are given in Table~\ref{tab:chandraobs}.  The merged
data set is of duration 93.568~ks.  For spectral analysis we created
appropriate response files for the separate observations and fitted
the four data sets to models in common.

To investigate the radio structure we use 1.375-GHz VLA
archival data from programme AC131 \citep{comins91}.  For the regions
around \nsix\ and \nseven\ we reduced the B-array data using AIPS.
Our image has a 4-arcsec FWHM restoring beam and a pixel size
matched to \chandra.  For the large-scale structure we have
used the smoothed version of the combined B, C and D-array image of
\citet{comins91} provided by J.P.~Leahy at
http://www.jb.man.ac.uk/atlas/, which has a 7.5-arcsec FWHM restoring
beam.  We have mapped the X-ray events directly
onto the radio grids and applied exposure corrections before combining
the images from the four \chandra\ observations when comparing the
radio and X-ray images.

\section{Morphology}
\label{sec:morph}

\begin{figure*}
\centering
%emulateapj%
\includegraphics[width=1.9\columnwidth,clip=true]{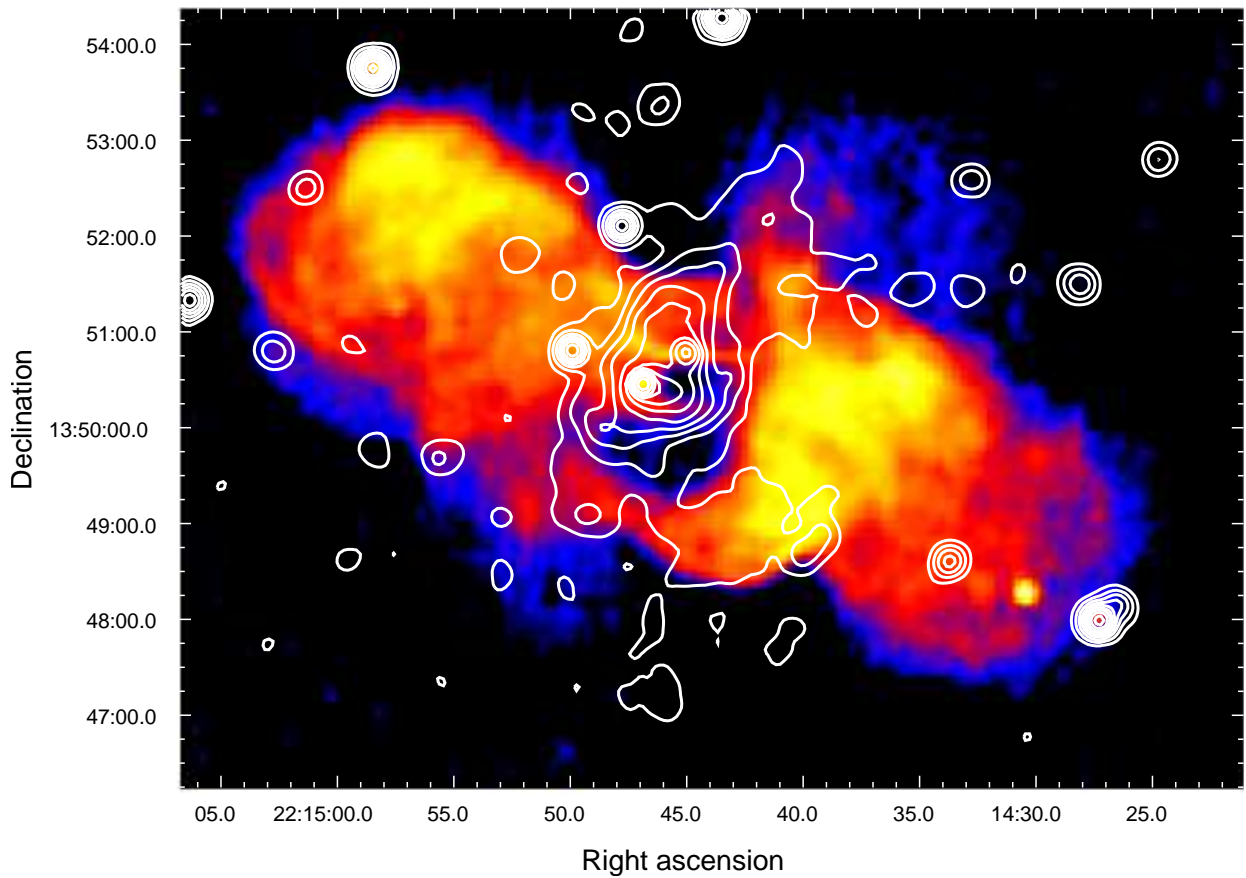}
\caption{X-ray contours (logarithmic spacing) on a colour image of the 1.4 GHz radio
emission.  The lowest contour is roughly at $3\sigma$ significance.
Emission from several
discrete sources is seen, including \nseven\ and \nsix.
Bright X-ray emission fills the gap between the lobes of \source.
\label{fig:XrayLimage}
}
\end{figure*}

\begin{figure}
\centering
%emulateapj%
\includegraphics[width=1.0\columnwidth,clip=true]{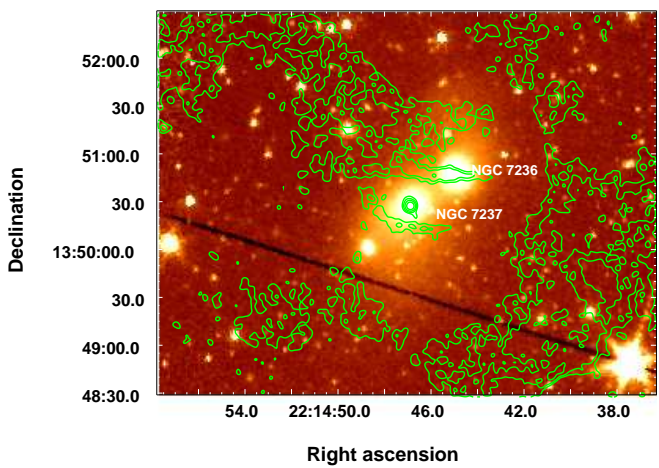}
\caption{\spitzer\ 4.5-$\mu$m image showing the diffuse red envelope
enshrouding \nsix\ and \nseven.  Contours (0.25, 0.5, 1, 2, 4, 8 mJy
beam$^{-1}$) are from the 1.4-GHz radio map with 4-arcsec FWHM
restoring beam, and highlight the relatively sharp inner edge of the
W lobe and the overall filamentary structure of the radio emission.
\label{fig:radioir}
}
\end{figure}

As seen from Figure~\ref{fig:XrayLimage}, the dominant X-ray emission
is diffuse, extending a few arcmin and elongated in a roughly NS
direction. The high spatial resolution and high sensitivity of the
\chandra\ data separates emission from \nsix\ and \nseven\, and a
third galaxy in the NW-SE chain (see Fig.~\ref{fig:radioir} for
their locations).
The inner X-ray emitting gas has a
distinctly higher surface brightness than larger-scale emission that
provides the background for the contours shown in
Figure~\ref{fig:XrayLimage}.

\begin{figure}
\centering
%emulateapj%
\includegraphics[width=0.7\columnwidth,clip=true]{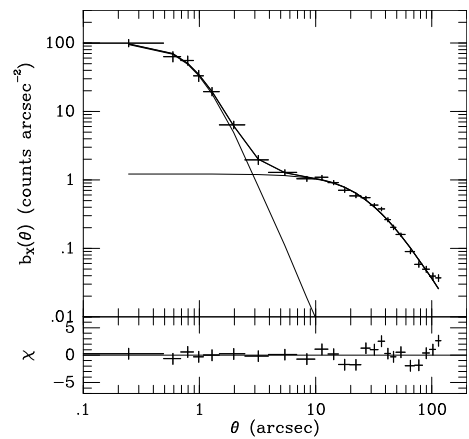}
\caption{0.3--5-keV radial profile centered on \nseven with other
discrete sources removed.  The fit to two $\beta$
models convolved with a nominal on-axis PSF 
gives $\chi^2 = 33.6$ for 20 degrees of freedom.  The relatively poor
fit for the larger-scale $\beta$ model, describing the gas between the
radio lobes, is indicative of an atmosphere that is not spherical and
may not have reached
dynamical or hydrostatic equilibrium. The
residuals at 100 arcsec point to a larger-scale,
lower-surface-brightness group gas that envelops the radio lobes
\citep[see][]{hard07}.
\label{fig:profile}
}
\end{figure}

Despite the elongated nature of the X-ray distribution, a radial
profile gives a good indication of the gas properties, and is
commonly used in the analysis of group and cluster X-ray emission.
We measure the component of most interest, the emission
between the radio lobes, by constructing a radial profile out to
2 arcmin from \nseven, masking other discrete sources and
using exposure-corrected background from a rectangular region between
3.6 and 5.7 arcmin to the NW which is observed in all four
observations.  The result for X-rays between 0.3 and 5 keV is shown in
Figure~\ref{fig:profile}.  \nseven\ is not precisely positioned at the
optical aim point in the four observations, giving different point
spread functions (PSFs).  For simplicity, we defer investigation of
\nseven\ \citep[see][]{hard07} and use an on-axis PSF convolved with a
$\beta$ model of very small core radius to fit the inner component.  The
outer component gives a reasonable fit to a convolved $\beta$ model
with best-fit values $\beta=0.62$, $\theta_{\rm c}= 28.4$ arcsec
(15.5~kpc), which is small for a normal group, and much less
than the scale of the group component on which this component is
superimposed \citep[Fig.~\ref{fig:profile} and][]{hard07}.

\section{Gas Properties}

The temperature of the bright gas has been measured by extracting
counts from a $\sim 1$~arcmin$^2$ rectangular box placed within the
outer contour of Figure~\ref{fig:XrayLimage}, and avoiding point
sources.  The background region is as described in \S\ref{sec:morph}.
The box was first placed north of \nsix, and then south of \nseven.
The data were fitted to an absorbed thermal model, and the parameter
values for the north and south locations were in good agreement.
No significant improvement in the fits were obtained when the absorption
was allowed to vary from the Galactic column density of $N_{\rm H} =
5.03 \times 10^{20}$ cm$^{-2}$ obtained from the {\sc colden} program
provided by the CXC and based on the data of \citet{dlock90}.  The
combined data from the two regions fit a thermal emission model of $kT
= 1.1^{+0.15}_{-0.1}$~keV with abundances $0.2^{+0.3}_{-0.1}$ of solar
($1\sigma$ uncertainties for 2 interesting parameters; $\chi^2 = 52$
for 51 degrees of freedom).

We use the equations of \citet{bw93} to deproject the $\beta$-model
X-ray surface-brightness profile (containing roughly 5000 counts out
to a radius of 90 arcsec) and combine it with our spectral results to
form a pressure distribution.  At a radius of 1 arcmin, roughly
corresponding to the position of inner contact between the gas and the
radio-lobe plasma, the pressure is $7.9^{+1.1}_{-1.7} \times 10^{-12}$
dynes cm$^{-2}$, where uncertainties are calculated as described in
\citet{wb01} and are dominated by the uncertainty in the abundances.
Our merger explanation for the gas (see below) makes it likely that
this is an underestimate of the total force felt by the lobes, because
the kinetic energy density of the gas has not yet entirely dissipated.
The total gas mass is $(3 \pm 1) \times 10^{10}$~M$_\odot$.

\section{Discussion}

While the component of group gas around \source\ has no unusual
characteristics,  the extra component of inner gas is remarkable.
We believe this gas is the result of the ongoing merger between
\nsix\ and \nseven, and was responsible for disrupting what was once a
rather ordinary radio galaxy emanating from either \nsix\ or \nseven.
The data support a picture in which, for about $10^8$ years
since the gas spheres overlapped, the lobes have been riding on the
pressure front of the merger gas which has been sweeping them apart.

If the reverse situation applied, and the
radio lobes were overpressured as compared with the inner gas, we
should expect to see morphological evidence for the influence of the
lobe on the gas, one example being a shock, as in the
SW inner lobe of Cen~A \citep{cenalobe}, although
here with the lobes expanding in towards the center.  That is not
seen.  Instead, the morphological appearance  is rather of the gas
separating the radio lobes, i.e., with the inner radio contours being
sharp and concave (rather than the normal lobe case of being convex),
as is particularly clear for the W lobe (Fig.~\ref{fig:radioir}).

Although the pressure of the gas touching the inner edges of the lobes
is
well measured by the \chandra\ data as $\sim 8 \times 10^{-12}$ dynes
cm$^{-2}$, the pressure inside the lobes, which on our interpretation
cannot be higher, is more uncertain.
Pressure estimates in radio lobes require assumptions about the
departure from particle and magnetic-field energy equipartition, filling
factors, and extrapolations of the observed radio spectrum to infer
the energy of the radiating (and non-radiating) particles.
\citet{comins91} estimate $4 \times 10^{-13}$ dynes cm$^{-2}$ for the
pressure assuming equal energy in electrons, protons, and magnetic
field (we agree). 
If the lobe pressure were indeed this low, then the lobes would be
under-pressured even with respect to the gas component at larger
($\gtrsim 2$~arcmin) radii, evident at the extremities of the radial
profile in Figure~\ref{fig:profile}, and so would be collapsing.
\citet{comins91} noted the filamentary nature of the radio structure
(Fig.~\ref{fig:radioir}) which they compared with the filamentation in
the Crab nebula, and suggested that the internal pressure should be
raised by a factor of 4--5 once the low filling factor is taken into
account.  There is much scope for raising the lobe pressure,
while still keeping it underpressured with respect to the inner gas.

In addition to the pressure-driven expansion of the gap between
the lobes, we expect old radio lobes to be of low density, and hence
buoyancy could also help to separate the lobes.
Since the pressure front at the edge of the merger gas will be moving
outwards at the sound speed, while
buoyant motion involves large-scale circulation that is only subsonic,
we expect the inner edges of the lobes to
be defined by the pressure front around the merger gas.
The morphology of the inner edge of the W radio lobe supports this
view, since it does not resemble a buoyant plume.

We suggest that
the merger between \nsix\ and \nseven\ created the $\sim 1$~keV gas envelope around the two
galaxies by heating the atmospheres of two gas-rich elliptical
galaxies.  The total gas mass of $\sim 3 \times 10^{10}$~M$_\odot$ is
rather high for the combined galaxy gas alone, and suggests a
contribution from group gas originally around one or both of \nsix\
and \nseven, as supported by the presence of the large-scale gas
component in the system \citep[see][]{hard07}.  We note,
however, that $ 10^{10}$~M$_\odot$ coronae have been
measured in isolated elliptical galaxies.  NGC~4555 is
such a case \citep{op04}, and shows sub-solar
abundances as we find in \source.  However, NGC~4555 
is a somewhat more massive galaxy than either major galaxy in \source, 
with a velocity dispersion of
$350\pm 1$ km s$^{-1}$ \citep{wegner}, as compared with $257\pm 17$ km
and $225\pm 31$ km s$^{-1}$ \citep{td81} for \nsix\ and \nseven,
respectively.  

\begin{figure}
\centering
%emulateapj%
\includegraphics[width=0.9\columnwidth,clip=true]{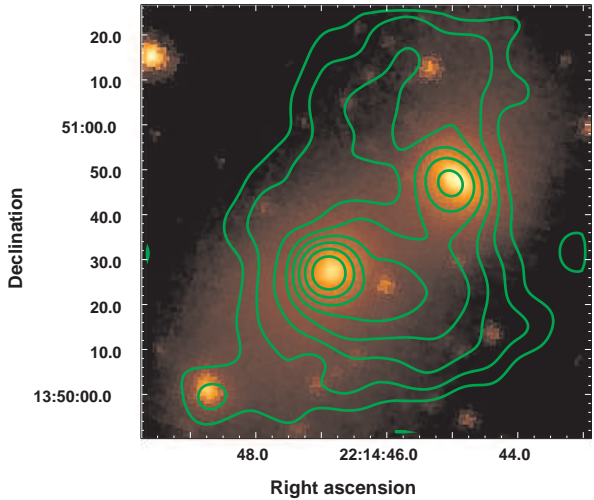}
\caption{Center of color image of Figure~\ref{fig:radioir} with X-ray contours of the
brightest emission.  The gas trails to the W of \nseven\ and NE of
\nsix\ are noticeably misaligned with the NW-SE merger fans seen in the infrared.
\label{fig:Xir}
}
\end{figure}

The sound-crossing time in the gas between the
lobes is $\sim 10^8$ years.  This would be the time scale on which the
radio lobes are being separated by the gas between them, and
it gives an approximate date for the violent phase of the merger that
would have heated the gas.
This is consistent with the prominent
large-scale stellar merger fans \citep[][Figs.~\ref{fig:radioir} \&
\ref{fig:Xir}, Birkinshaw et al., in preparation]{borneh88}
which are typically visible for $\sim 10^8$ years \citep{vandokkum}.

Partial ram-pressure stripping of gas in ellipticals due to their
motion within a cluster or group atmosphere is commonly reported
\citep[e.g.,][]{sun}.  In \source\ the properties of the inner gas
suggest a more violent galaxy-galaxy encounter.  In the modelling of
\citet{borne88} the galaxy centres passed within 8 kpc of one another
about $10^8$ years ago, so their gas
spheroids would have interacted strongly.  In such an encounter, most of the gas
in the two galaxies would have been extracted and added to the center
of the group after the heating that would have resulted from the
dissipation of kinetic energy.  This would have expanded into, and mixed with, the
hotter ambient group gas.  The gas trails behind the galaxies, seen in
Fig~\ref{fig:Xir} and discussed more fully in \citet{hard07}, lend
support to this picture.  As the new gas-component expands, it pushes the 
lobes apart.  In this picture, the radio
filaments lying within the gas (Fig.~\ref{fig:radioir}) are small
pieces of the radio lobes (or extinct jets) that have been engulfed
and crushed by the central gas. 

The internal energy of the gas between the lobes is $\sim 5 \times
10^{58}$ ergs, which is more than the energy radiated by the radio
source since the merger \citep[$\sim 3 \times 10^{56}$ ergs in $10^8$
years:][]{comins91}, and $\sim 10$\% of our estimate of the energy
in the lobes, and so it is likely that some of the lobe brightness,
particularly in the inner W lobe, is due to compression caused by the
central gas.  The X-ray gas is thus a source of energy that may
re-excite electrons in the lobes through adiabatic heating, and which
will raise the
radio brightness of the lobes.
Similar compressions of radio-emitting plasma may be
responsible for re-energizing particles in cluster radio relic
sources.  
The inner radio lobes of \source\ may be sites where
this process can be studied closely in a case, like some radio relics,
where strong shocks are not apparent
\citep[e.g.,][]{feretti}.

In summary, the inner gas of \source\ is a distinct component, which
can be separated from ambient larger-scale group gas.  
Our view is that the gas of \nsix\
and \nseven\ has merged as a result of a close encounter, and that the
dissipation of kinetic energy has heated this gas at the
same time as it mixes with the ambient (hotter) group medium.  This
central gas is overpressured with respect to the radio lobes whose jets
may have ceased at about the time of, and possibly as a result of, the
merger.  The pressure-driven expansion of the merger-liberated gas is
compressing and separating the lobes,
and may be re-exciting electrons in the lobe plasma
through adiabatic heating.  

\acknowledgments

We thank the CXC for its support of \chandra\ observations,
calibrations, data processing and analysis, the SAO R\&D group for
{\sc DS9} and {\sc funtools}, and Alexey Vikhlinin and Eric Mandel for
additional software used for part of our analysis. MJH thanks the
Royal Society for a University Research Fellowship.
This work has used data from the VLA.
NRAO is a facility of the National Science Foundation operated under
cooperative agreement by Associated Universities, Inc.
The work has been partially supported by NASA grant GO5-6100X.


\begin{thebibliography}{}
\bibitem[Birkinshaw \& Worrall(1993)]{bw93} Birkinshaw, M., \& Worrall
D.M.~1993, \apj, 412, 568
\bibitem[Birkinshaw, Laing \& Peacock(1981)]{birk81} Birkinshaw, M.,
Laing, R.A., \& Peacock, J.A.~1981, \mnras, 197, 253
\bibitem[Borne(1988)]{borne88} Borne, K.D.~1988, \apj, 330, 61
\bibitem[Borne \& Hoessel(1988)]{borneh88} Borne, K.D., \& Hoessel J.G.~1988, \apj, 330, 51
\bibitem[Comins \& Owen(1991)]{comins91} Comins, N.F., \& Owen F.N.~1991, \apj, 382, 108
\bibitem[Dickey \& Lockman(1990)]{dlock90} Dickey,
  J.M., \& Lockman, F.J.~1990, ARA\&A, 28, 215
\bibitem[Feretti \& Neumann(2006)]{feretti} Feretti, L., \& Neumann, D.M.~2006,
\aa, 450, L21
\bibitem[Hardcastle \& Worrall(1999)]{hw99} Hardcastle M.J., \& Worrall
D.M.~1999, \mnras, 309, 969
\bibitem[Hardcastle et al.(2007)]{hard07} Hardcastle M.J., Kraft R.P.,
Worrall, D.M., Croston, J.H., Evans, D.A., Birkinshaw, M., \& Murray,
S.S.~2007, \apj, submitted
\bibitem[Jones et al.(2007)]{jones07} Jones, C., Forman, W.,
Vikhlinin, A., Markevitch, M., Machacek, M., \& Churazov, E.,  2007,
in {\it A Pan-Chromatic View of Clusters of Galaxies and the
Large-Scale Structure}, ed. M.~Plionis \& O.~Lopez-Cruz
(Berlin: Springer Verlag), in press
\bibitem[Kraft et al.(2003)]{cenalobe} Kraft, R.P., V\'azquez, S.,
Forman, W.R., Jones, C., Murray, S.S.,
Hardcastle, M.J., Worrall, D.M., \& Churazov, E.~2003, \apj, 592, 129
\bibitem[O'Sullivan \& Ponman(2004)]{op04} O'Sullivan, E., \& Ponman,
T.J.~2004, \mnras, 354, 935
\bibitem[Sun et al.(2007)]{sun} Sun, M., Jones, C., Forman, W.,
Vikhlinin, A., Donahue, M., \& Voit, M.~2007, \apj, in press
\bibitem[Tonry \& Davis(1981)]{td81} Tonry, J.I., \& Davis, M.~1981,
\apj, 246, 666
\bibitem[van Dokkum(2005)]{vandokkum} van Dokkum, P.G.~2005, \aj,
130, 2647
\bibitem[Wegner et al.(2003)]{wegner} Wegner, G. et al.~2003, \aj,
126, 2680
\bibitem[Worrall \&  Birkinshaw(2001)]{wb01} Worrall
D.M., \& Birkinshaw, M.~2001, \apj, 551, 178
\end{thebibliography}
\end{document}